\documentclass{iau}
\usepackage{graphicx,natbib}
 
\newcommand{\apjl}{ApJ}           
\newcommand{\mnras}{MNRAS}       

\newcommand{\aap}{A\&A}

\newcommand\ageL{$\langle {\rm log}\,age\rangle _L$}

\newcommand\HLR{$a_{50}^L$}

\newcommand\logZM{$\langle  \log Z_\star \rangle _M$}

\title{The star formation history of galaxies in 3D: The CALIFA perspective}

\author[Gonz\'alez Delgado et al]{R.M. Gonz\'alez Delgado$^1$, R. Cid Fernandes$^2$, R. Garc\'\i a-Benito$^1$, E. P\'erez$^1$, A. L. de Amorim$^2$, C. Cortijo-Ferrero$^1$, E. A. D. Lacerda$^2$,  R. L\'opez Fern\'andez$^1$, S. F. S\'anchez$^{1, 3}$, N. Vale Asari$^2$,   \and CALIFA collaboration}

\affiliation{$^1$Instituto de Astrof\'\i sica de Andaluc\'\i a (CSIC), Glorieta de la Astronom\'\i a s/n, E-18008 Granada, Spain \\
email: {\tt rosa@iaa.es} \\
$^2$Departamento de F\'\i sica, Universidade Federal de Santa Catarina, P.O. Box 476, 88040-900, Florian\'opolis, SC, Brazil \\
$^3$ Instituto de Astronom\'\i a,Universidad Nacional Auton\'oma
de Mexico, A.P. 70-264, 04510, M\'exico,D.F.}
\pubyear{2014}
\volume{309}
\jname{Galaxies in 3D across the Universe}
\editors{B. L. Ziegler, F. Combes, H. Dannerbauer, M. Verdugo, eds.}

\begin{document}

\maketitle

\begin{abstract}
We spatially resolve  the star formation history of 300 nearby galaxies from the CALIFA integral field survey to investigate: a) the radial structure and gradients of the present stellar populations properties as a function of the Hubble type; and b) the role played by the galaxy's stellar mass and stellar mass surface density in governing the star formation history and metallicity enrichment of spheroids and disks. We apply the fossil record method based on spectral synthesis techniques to recover spatially and temporally resolved maps of stellar population properties of spheroid and spiral galaxies with mass from $10^9$ to $7 \times 10^{11}$ M$_\odot$. The individual radial profiles of the stellar mass surface density ($\mu_\star$), stellar extinction (A$_V$), luminosity weighted ages (\ageL), and mass weighted metallicity (\logZM) are stacked in seven bins of  galaxy morphology (E, S0, Sa, Sb, Sbc, Sc and Sd). All these properties show negative gradients as a sign of the inside-out growth of massive galaxies. However, the gradients depend on the Hubble type in different ways. For the same galaxy mass, E and S0 galaxies show the largest inner gradients in $\mu_\star$; while Andromeda-like galaxies (Sb with M$_\star \sim 10^{11} $ M$_\odot$)  show the largest inner age and metallicity gradients. On average, spiral galaxies have a stellar metallicity gradient $\sim$ -0.1 dex per half-light radius, in agreement with the value estimated for the ionized gas oxygen abundance gradient by CALIFA.  Both global (M$_\star$-driven) and local ($\mu_\star$-driven) stellar metallicity relation are derived. We find that in disks, the stellar mass surface density regulates the stellar metallicity; in spheroids, the galaxy stellar mass dominates the physics of star formation and chemical enrichment.

\end{abstract}

\firstsection
\section{Introduction}

Much of we know about galaxy properties has come from panoramic imaging or 1D spectroscopic surveys. While imaging provides useful 2D information of galaxy morphology and structural properties, spectroscopic surveys give information on the central  or global stellar population and ionized gas properties and kinematics. However, galaxies are a complex mix of stars, interstellar gas, dust and dark matter, distributed in their disks and bulges, and resolved spatial information is needed to constrain the formation processes and evolution of the galaxy sub-components.  

Integral Field Spectroscopy (IFS) observations can provide a unique 3D view of galaxies (two spatial plus one spectral dimensions), and allows to recover 2D maps of the stellar population and ionized gas properties. Until very recently it was not possible to obtain IFS data for sample larger than a few tens of galaxies. The ATLAS3D (Cappellari et al.\ 2011), CALIFA (S\'anchez et al.\ 2012), SAMI (Croom et al.\ 2012) and MaNGA (Bundy et al.\ 2014) surveys are the first in doing this step forward, taking observations of several hundreds to several thousands of galaxies of the nearby Universe.

CALIFA is our currently-ongoing survey, observing 600 nearby galaxies with the PPaK IFU at the 3.5m telescope of Calar Alto observatory. The observations cover 3700--7000 \AA\ with an intermediate spectral resolution (FWHM $\sim$ 6 \AA\ in the data presented in this contribution) and $\sim$1arcmin$^2$ field of view with a final spatial sampling of 1 arcsec. Galaxies were selected from SDSS in the redshift range 0.005 $\leq$ z $\leq$ 0.03, covering all the color magnitude diagram down to M$_r\leq$ -18, resulting in a sample containing all morphological types. An extended description of the survey, data reduction and sample can be found in S\'anchez et al (2012), Hussemann et al (2012) and Walcher et al (2014).

Previously, we derived the spatially resolved star formation history of the CALIFA galaxies on the first data release (DR1) 
using the fossil record of stellar populations imprinted in their spectra. This method dissects galaxies in space and time providing a 3D information that allows to retrieve when and where the mass and stellar metallicity were assembled as a function of look-back-time. We use the {\sc starlight} code (Cid Fernandes et al.\ 2005) to do a  $\lambda$-by-$\lambda$ spectral fit using different sets of single stellar population (SSP) models. These SSPs are from a combination of Vazdekis et al.\ (2010) and Gonz\'alez Delgado et al.\ (2005) (labelled {\it GMe}), or from Charlot \& Bruzual (2007) (labelled {\it CBe}).

Our scientific results from the first 100 CALIFA galaxies were presented in P\'erez et al (2013),
Cid Fernandes et al (2013, 2014) and Gonz\'ales Delgado et al (2014a). One highlight result of these works is that the signal of downsizing is spatially preserved, with inner and outer regions growing faster for more massive galaxies, consequence of the inside-out growth of massive galaxies.

Here, based on the analysis of 300 CALIFA galaxies we present the results on: a) the radial structure and gradients of the stellar populations as a function of the Hubble type; and b) the roles of the galaxy mass and stellar mass surface density in governing the star formation history and metallicity enrichment in ellipticals and in the bulge and disk components of galaxies.

Two complementary contributions to this one are presented in these proceedings: S\'anchez presents CALIFA in the context of other contemporaneous IFS surveys such as  SAMI and MaNGA, and Cid Fernandes et al.\ highlights some previous and new results obtained with the same methodology applied here.

\section{Results}

We present new results based on the  radial structure of the present stellar population properties of 300 CALIFA galaxies that were observed with the V500 and V1200 setups and calibrated with the new pipeline 1.4 (see Garc\'\i a-Benito et al.\ 2014 for details). 1D radial profiles are obtained from the 2D maps, with an azimuthal averaging by an elliptical $xy$-to-$R$ conversion. These 2D maps are created after collapsing the star formation history (SFH)  in the time domain. The results are presented by stacking each galaxy individual radial profiles after normalizing to a common metric that uses the half-light-radius (HLR) of each galaxy.

\subsection{Hubble sequence: stellar population properties of galaxies in the tuning-fork diagram}

A  step to understand how galaxies form and evolve is classifying galaxies and studying  their properties. Most of the massive galaxies in the near Universe are E, S0 and spirals (Blanton \& Moustakas 2009), following well the Hubble tuning-fork diagram.  The bulge fraction seems to be one of the main physical parameters that produce the Hubble sequence, increasing from late to early spirals. In this scheme, S0 galaxies are a transition class between the spiral classes and the elliptical one, with large bulges, but intermediate between Sa and E galaxies. On the other hand,  properties such as color, mass, surface brightness, luminosity, and gas fraction are correlated with the Hubble type (Robert \& Haynes 1994). This suggests that the Hubble sequence can illustrate possible paths for galaxy formation and evolution. If so, how are  the spatially resolved stellar population properties of galaxies correlated with the Hubble type? Can the Hubble-tuning-fork scheme be useful to organize galaxies by galaxy mass and age or  galaxy mass and metallicity? 

CALIFA is a suitable benchmark to address these questions because it includes a significant amount of E, S0 and spirals. After a visual classification, the 300 galaxies were grouped in 41 E,  32 S0, 51 Sa, 53 Sb, 58 Sbc, 50 Sc, and 15 Sd. This sub-sample is representative of the morphological  distribution of the whole CALIFA sample. Here we present the radial structure of the stellar mass surface density ($\mu_\star$), stellar extinction (A$_V$), luminosity weighted stellar age (\ageL),  mass weighted stellar metallicity (\logZM), by stacking the galaxies by their Hubble type. First, we present how galaxies are distributed by stellar mass (M$_\star$) and their sizes in mass ($a_{50}^M$: radius that contains half of the mass) and in light ($a_{50}^L$: radius that contains half of the light, HLR). Most of the results discussed here are obtained with the  {\it GMe} SSP models, but similar results are obtained with the {\it CBe} base (see Fig.\ 1). 

{\bf Galaxy stellar mass:}
We obtain the galaxy stellar mass (M$_\star$) after resolving spatially the SFH of each zone, therefore  taking into account spatial variations of the stellar extinction and M/L ratio. Fig.\ 1a shows the distribution of M$_\star$ as a function of Hubble type. The mass ranges from 10$^9$ to 7$\times$10$^{11}$ M$_\odot$ (for {\it GMe} SSPs). We see a clear segregation in mass: galaxies with high bulge-to-disk ratio (E, S0, Sa) are the most massive ones ($\geq$ 10$^{11}$ M$_\odot$), and galaxies with small bulges (Sc-Sd) have masses M $\leq$ 10$^{10}$ M$_\odot$. The stellar mass distribution obtained with {\it CBe} models is similar to that obtained  with the {\it GMe} base, but shifted by $-0.25$ dex due to the difference in IMF (Chabrier in {\it CBe} and Salpeter in {\it GMe}). 

{\bf Galaxy size:} We take advantage of our spatially resolved SFH and extinction maps to show that 
 galaxies are more compact in mass than in light (Gonz\'alez Delgado et al.\ 2014a), resulting in a ratio of the radius that contains half of the mass ($a_{50}^M$) with respect to the radius that contains half of the light ($a_{50}^L$) of  $a_{50}^M/a_{50}^L \sim 0.8$. Galaxies are therefore typically $20 \%$ smaller in mass than  they appear in optical light. Fig.\ 1b shows the distribution of $a_{50}^M/a_{50}^L$ as a function of Hubble type, and  a clear trend is observed. Sa-Sb-Sbc have the lowest $a_{50}^M/a_{50}^L$ ratios, due to the fact that they show a very prominent old bulge which have similar central properties to the spheroidal components of S0 and E, but a blue and extended disc which contributes to the light in the range Sa to Sbc. 
 
\begin{figure}
\centering
\includegraphics[width=0.49\columnwidth]{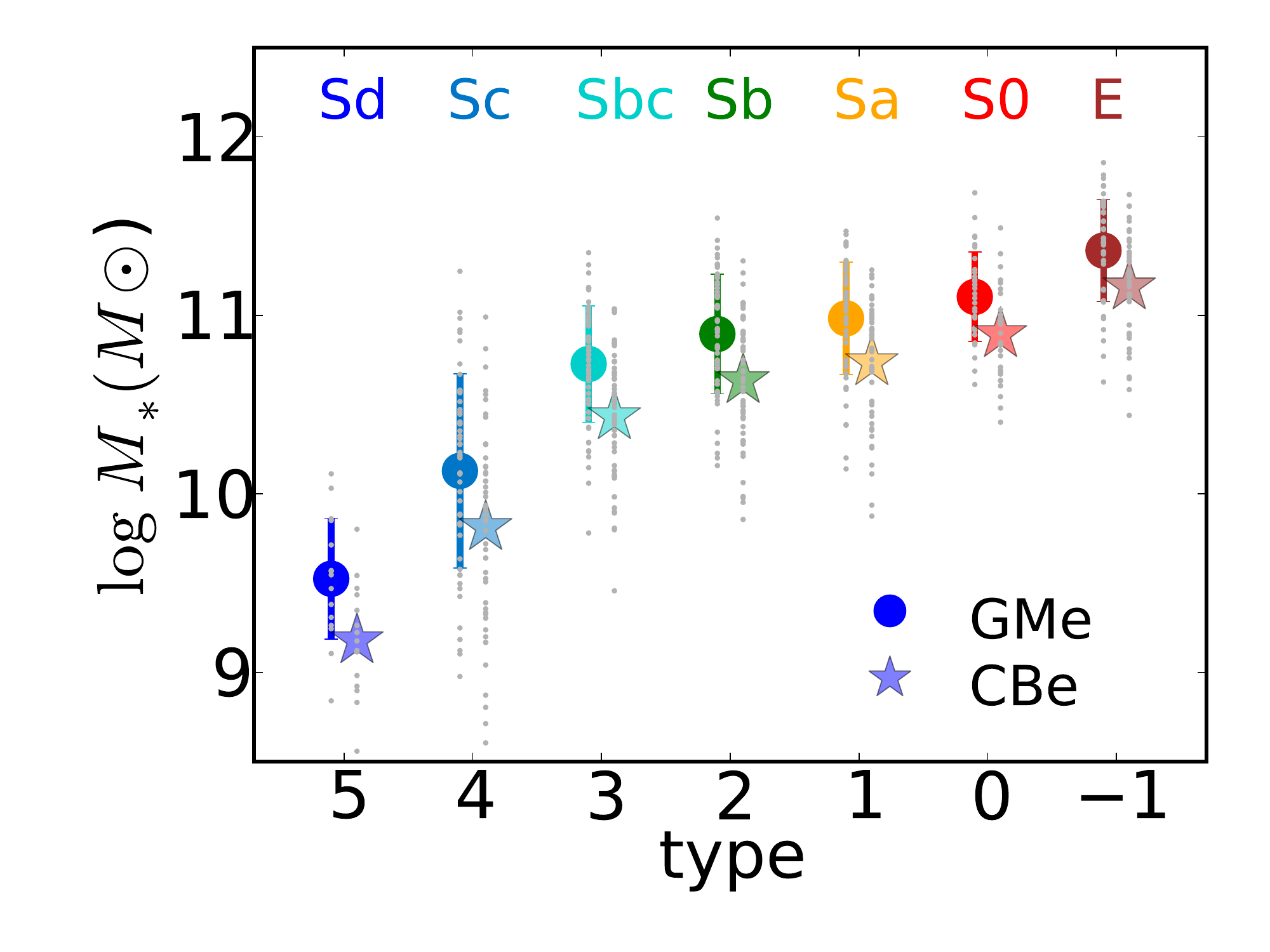} 
\includegraphics[width=0.49\columnwidth]{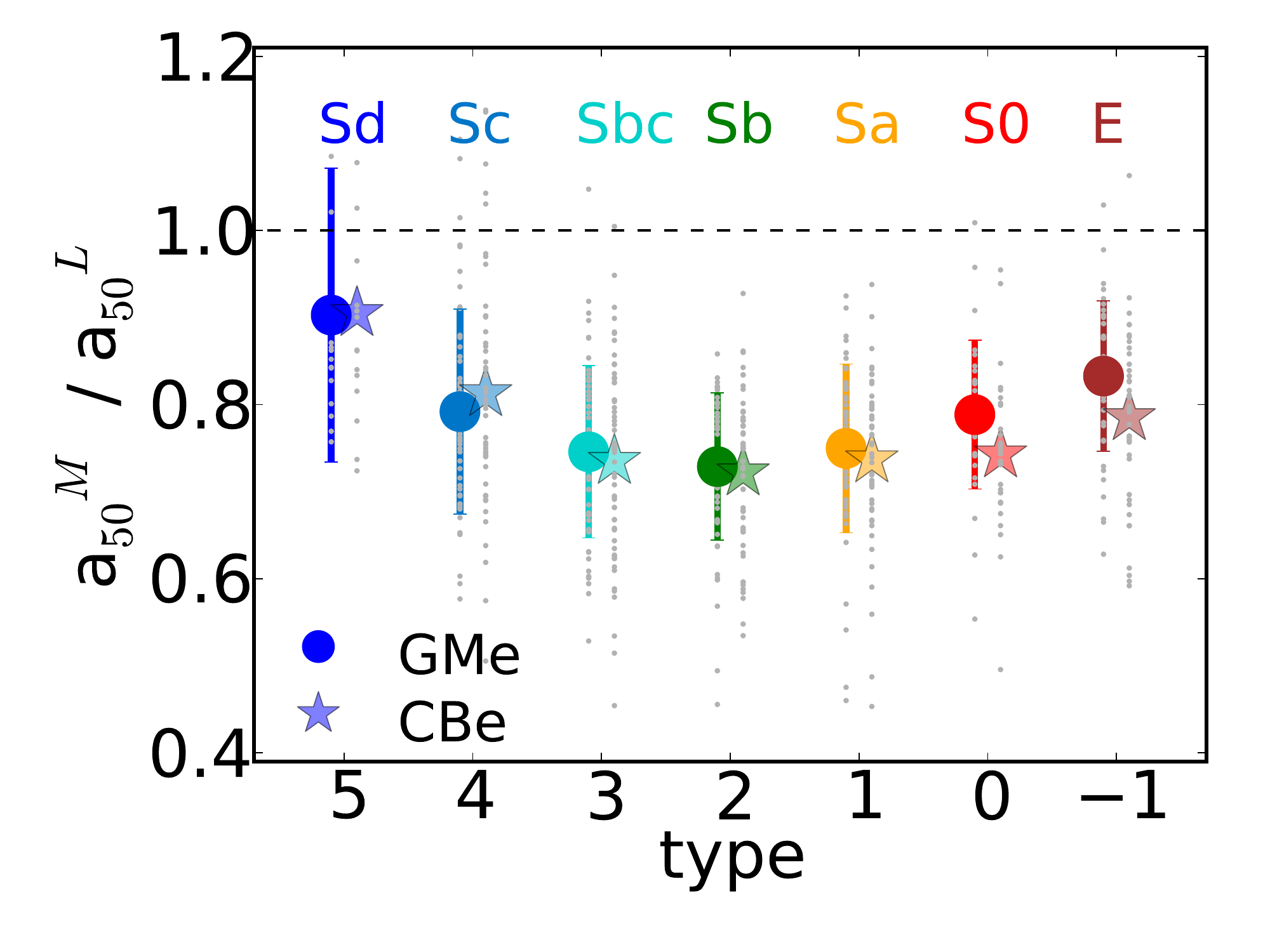} 
\caption{Left: Distribution of the galaxy stellar masses obtained from the spatially resolved spectral fits of each galaxy for each Hubble type of the galaxy of this work (grey small points). The coloured dots and stars are the mean galaxy stellar mass in each Hubble type obtained with the GMe and CBe SSP models. The bars show the dispersion in mass. Right: Relation between $a_{50}^M/a_{50}^L$ and Hubble type. Symbols are as in the left panel of the figure.}
\label{fig:fig1}
\end{figure}

{\bf Stellar mass surface density:}
The left panel in Fig.\ 2 shows the radial profiles (in units of a$_{50}^L$) of $\log\ \mu_\star$, as obtained with the {\it GMe} base. Individual results are stacked in seven morphological bins. Error bars in the panel indicate the dispersion at one a$_{50}^L$ distance in the galaxies of the Sa class, but it is similar for other Hubble types and radial distances.  Negative gradients are detected in all galaxy types and increase from late type (Sd) to early type (S0, and E) galaxies. At a constant M$_\star$, spheroidal (S0 and E) are more compact than spirals, and  S0 and E galaxies have  similar compactness at all distances. 

{\bf Stellar extinction:}
The right panel in Fig.\ 2 shows the radial profiles A$_V$. All galaxy types show radial profiles that increase towards the center. Early type galaxies (E, and S0) are also extinguished towards the nucleus by 0.2 mag, while outwards of 1 HLR their (old) stellar populations are almost reddening-free. All spiral disks show $\sim 0.2$--0.3 mag extinction. The bulges are significantly more extinguished up to 0.6 mag, except in late type spirals (Sd), where $A_V$ is similar to the extinction in the disk.
We note that , in absolute terms, one may thus say that galaxies become dustier towards their centers. In relative terms, however, the opposite holds. The typical $A_V/\mu_\star$ profiles rise towards the outside, so one can say that there is more dust per star at large radii, which indirectly signals also a rise in the gas-to-stars at large $R$.

\begin{figure}
\centering
\includegraphics[width=0.49\columnwidth]{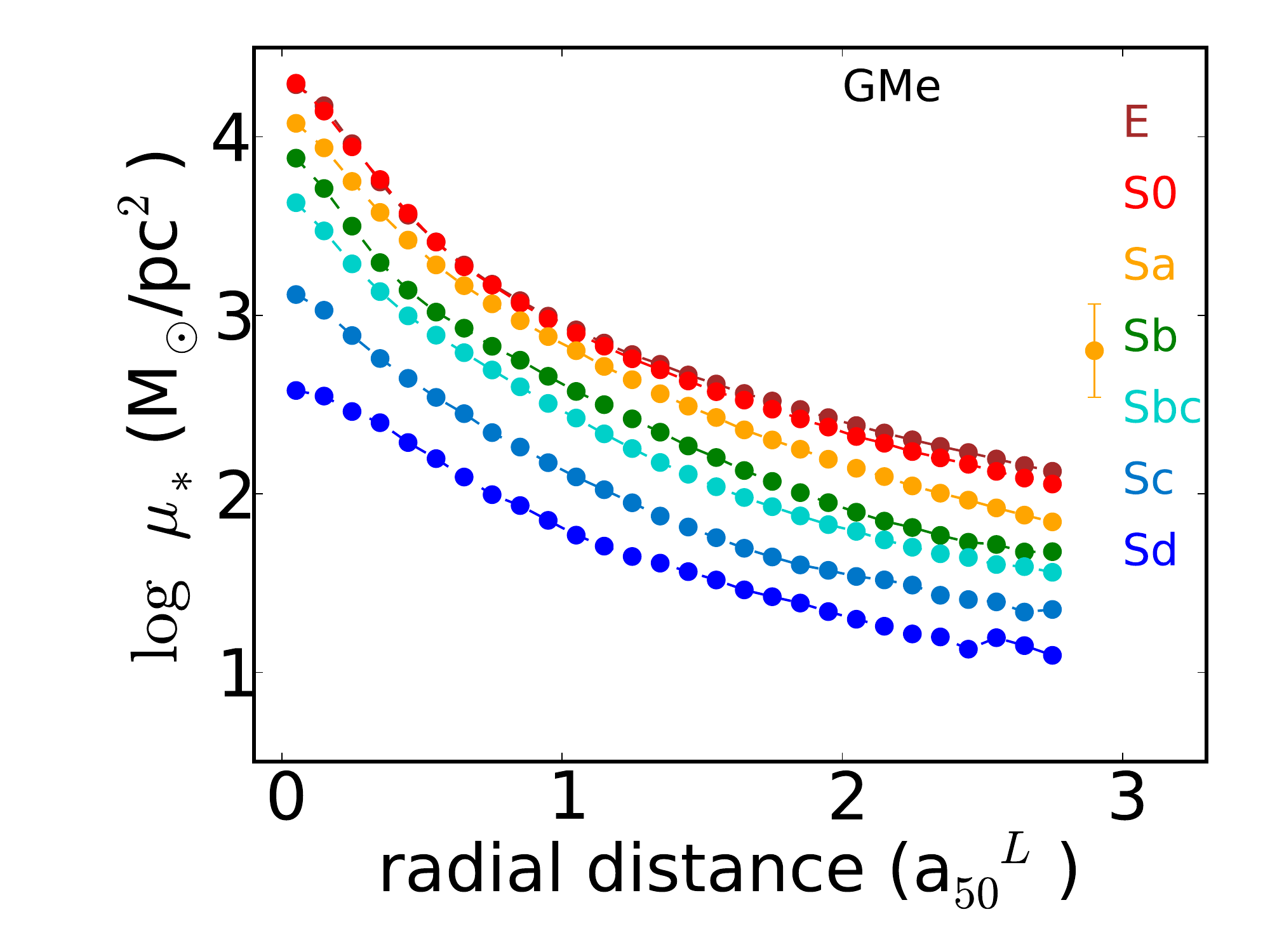} 
\includegraphics[width=0.49\columnwidth]{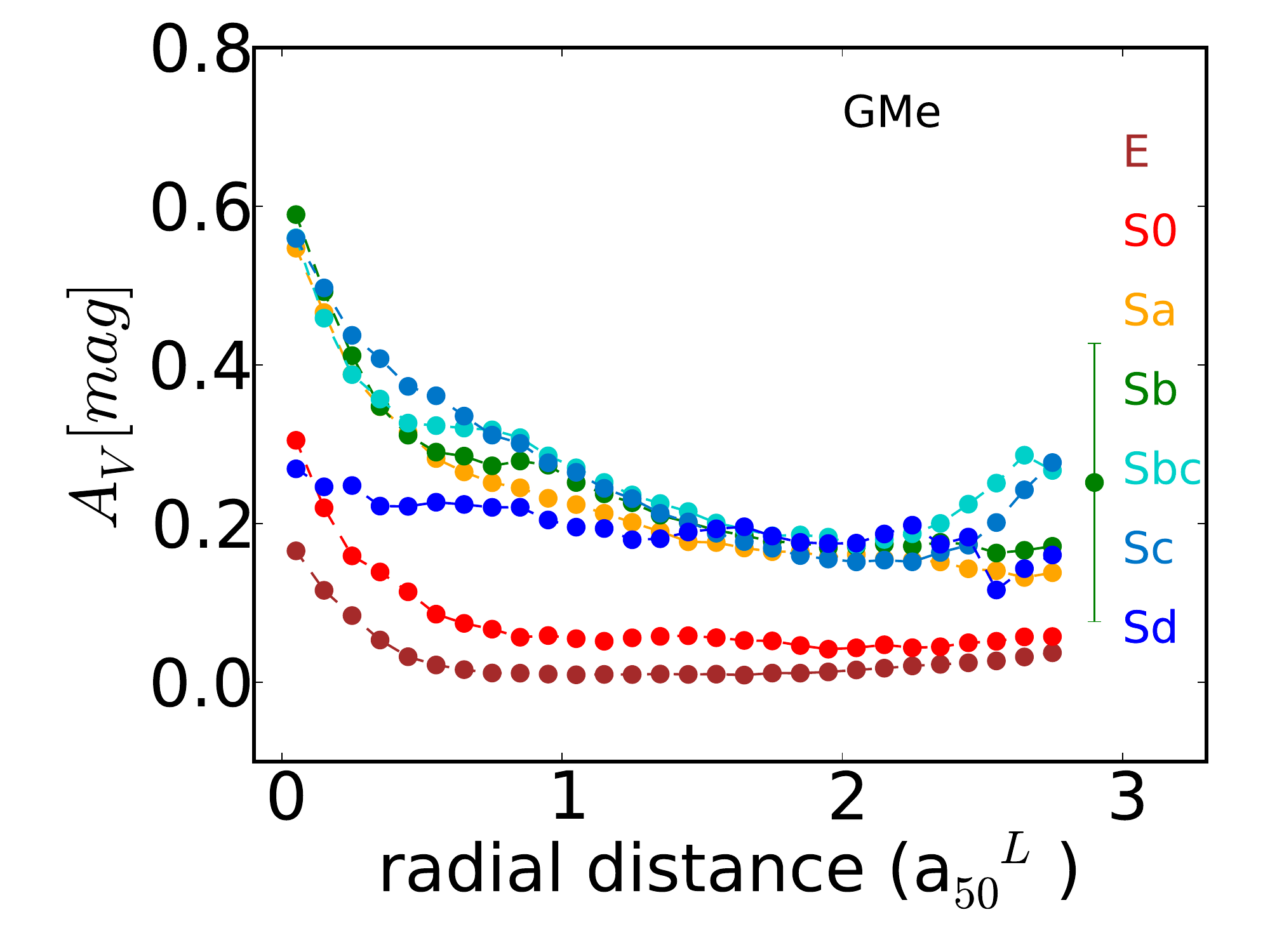} 
\caption{Radial profiles of the stellar mass surface density (in logarithm scale, $\log\ \mu_\star$) and the stellar extinction A$_V$ as a function of Hubble type.  The radial distance is  in HLR units.}
\label{fig:fig2}
\end{figure}

{\bf Stellar ages:}
The left panel of Fig.\ 3 shows the radial profiles of  \ageL. Symbols are as in Fig.\ 2. Negative gradients are detected for all the Hubble types, suggesting that the quenching is progressing outwards, and the galaxies are growing inside-out, as we concluded with  our mass assembly growth analysis (P\'erez et al.\ 2012). Inner gradients are calculated between the galaxy nucleus and at 1 \HLR, and the outer gradient between 1 and 2 \HLR. The inner age gradient shows a clear behaviour with Hubble type, being maximum for spirals of intermediate  type (Sb-Sbc). At constant M$_\star$, Sb-Sbc galaxies have the largest age gradient. The age gradient in the outer disk (between 1 and 2 \HLR) is smaller than the inner ones, but again it is largest amongst Sa-Sb-Sbc's. 

\begin{figure}
\centering
\includegraphics[width=0.49\columnwidth]{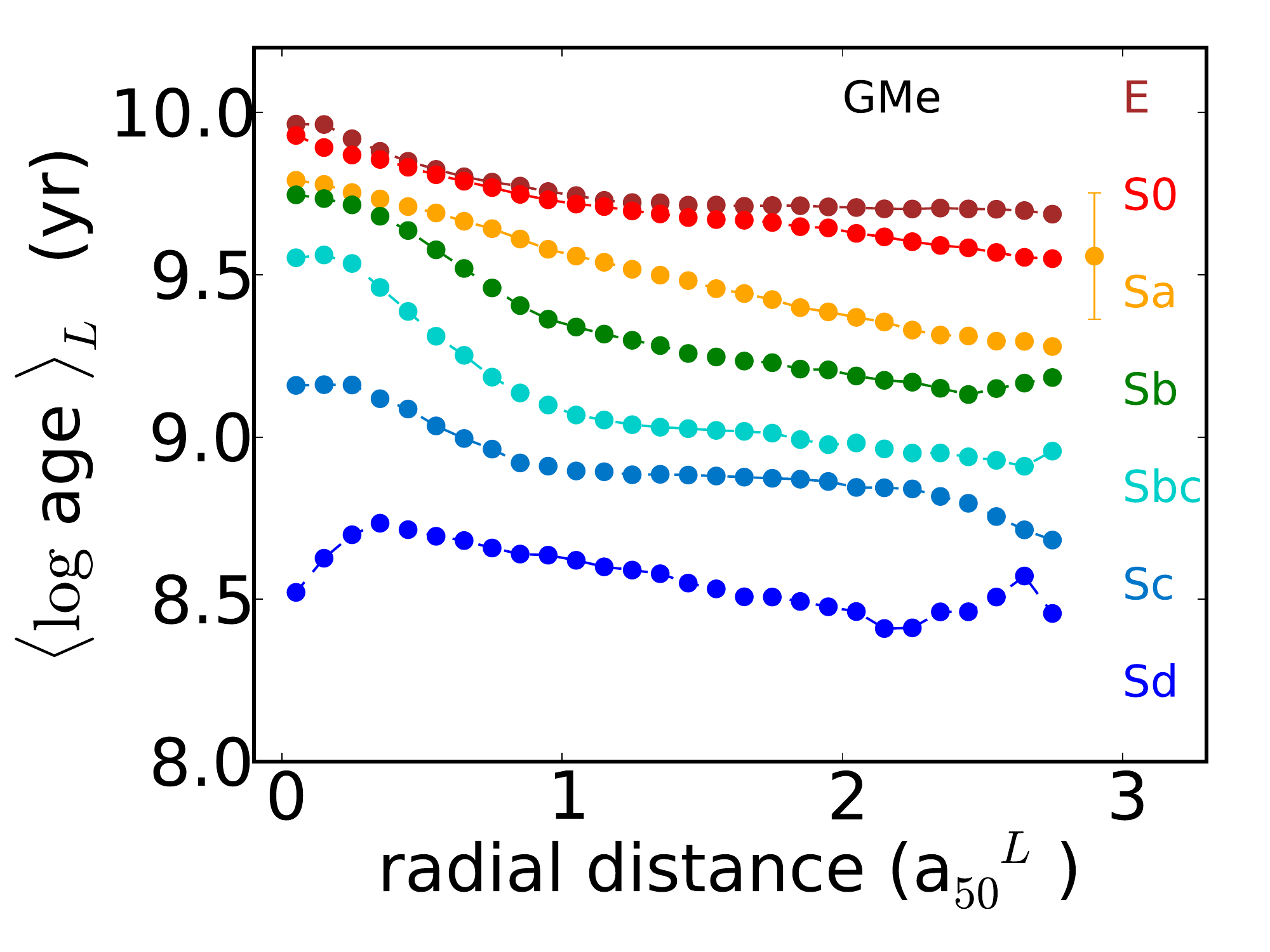}
\includegraphics[width=0.49\columnwidth]{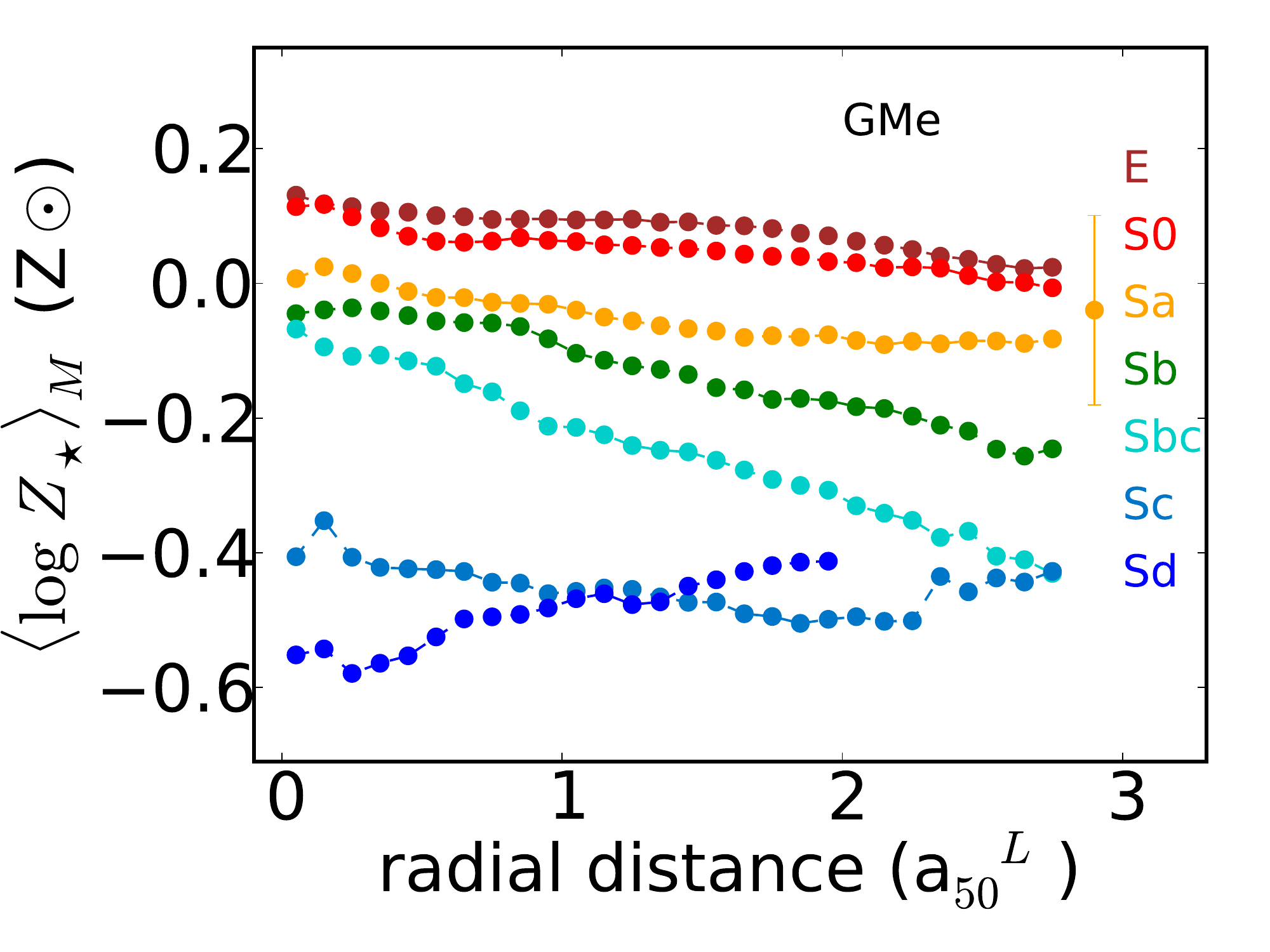}  
\caption{As in Fig.\ 2 for the luminosity weighted ages and the mass weighted metallicity.}
\label{fig:fig3}
\end{figure}

{\bf Stellar metallicity:}
The right panel in Fig.\ 3 shows the radial profiles of mass weighted stellar metallicity obtained as explained in Gonz\'alez Delgado et al (2014b). Except for late types (Sc-Sd), spirals present negative $Z_\star$ gradients: On  average $\sim -0.1$ dex per HLR, similar to the value obtained for the nebular oxygen abundances obtained by CALIFA (S\'anchez et al.\ 2013). For galaxies of equal stellar mass, the  intermediate type spirals (Sbc) have the largest gradients. These negative gradients again are signs of the inside-out growth of the disks. However, the metallicity gradient for late spirals is positive indicating that these galaxies, in the low mass galaxy bins, may be formed from outside-in. This result is also in agreement with our results from the mass assembly evolution (P\'erez et al.\ 2012).

\subsection{The role of $\mu_\star$ and M$_\star$ in the SFH and chemical enrichment of galaxies}

Galaxy formation and evolution manifests in three main parameters: stellar age, mass and metallicity. 
The color-magnitude diagram of galaxies --that charts our partial knowledge of galaxy evolution in the bimodal distribution
of red sequence and blue cloud-- is the observational proxy for the two fundamental evolutionary parameters: 
the age and mass of the stellar populations in galaxies.
Sorting galaxies by M$_\star$ we can study how their properties scale among the different classes. With CALIFA, thanks to the spatial information, we can check how important are the local ($\mu_\star$- driven) and global (M$_\star$-driven) processes in determining the star formation history and chemical enrichment in galaxies.

We find that there is a strong relation between the local values of $\mu_\star$ and the metallicity, a relation which is similar in amplitude to the global mass metallicity relation that exits between \logZM\ and M$_\star$ over the whole $10^9$ to $10^{12}$ M$_\odot$ range (Gonz\'alez Delgado et al.\ 2014b). This means that local and global processes are important in the metallicity enrichment of the galaxies. However, the balance between local and global effects varies with the location within a galaxy. While in disks, $\mu_\star$ regulate the stellar metallicity, producing a  correlation between  $\log\ \mu_\star$ and \logZM, in bulges and ellipticals is M$_\star$ who dominates the chemical enrichment (Fig.\ 4). Furthermore, in  spheroids the chemical enrichment happened much faster and earlier than in disks.

These results are in agreement with the analysis of the star formation history of galaxies (Gonz\'alez Delgado et al.\ 2014a). We have shown that mean stellar ages (a first moment descriptor of the SFH) relate strongly to $\mu_\star$ in galactic disks, indicating that local properties dictate the pace of star-formation. The slower growth (hence younger ages) found at low $\mu_\star$ should lead to less metal enrichment, in agreement with the $\mu_\star$-$Z_\star$ relation depicted in Fig.\ 4b. Within bulges/spheroids, $M_\star$ is a much more relevant driver of the SFH. Most of the star formation activity in these regions was over long ago, leading to fast metal enrichment and little or no chemical evolution since those early days.

\begin{figure}
\centering
\includegraphics[width=\columnwidth]{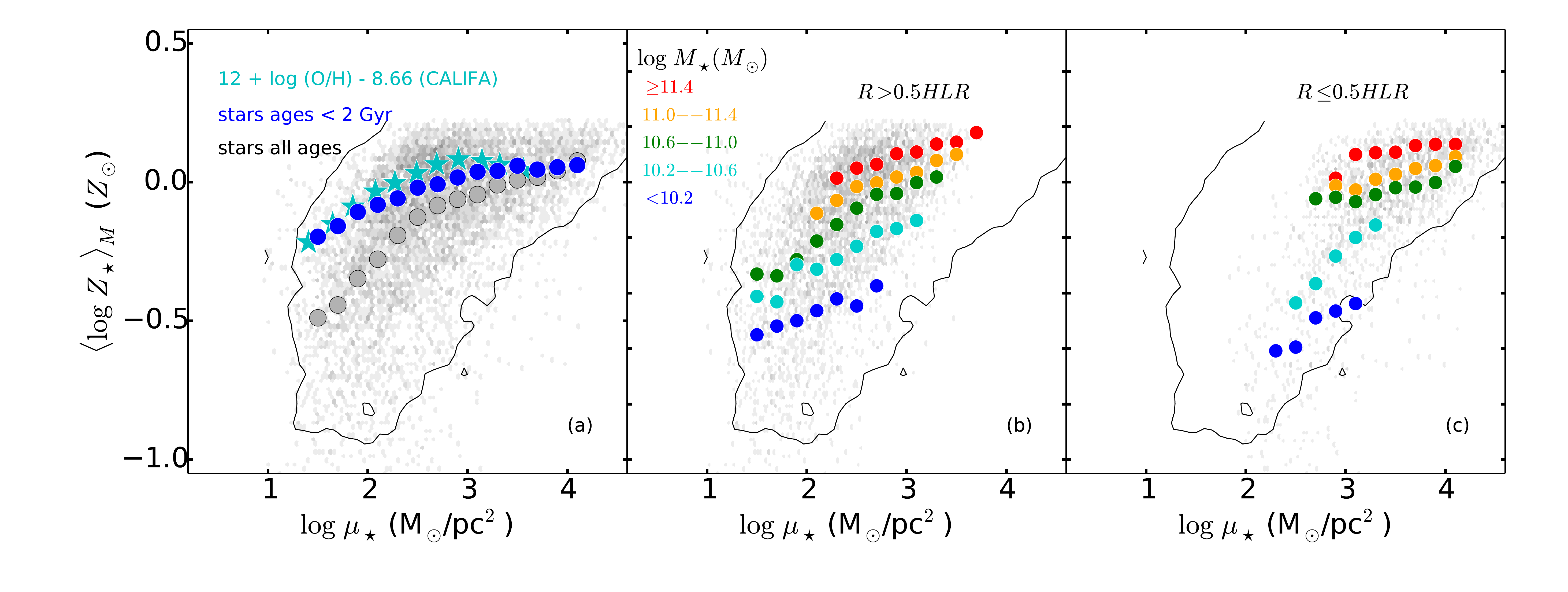} 
\caption{Left: Local stellar metallicity versus the local stellar mass surface density. The grey circles tracks the $\mu_\star$-binned stellar metallicity relation ($\mu$ZR). Blue circles show the $\mu$ZR obtained considering only star younger than 2 Gyr  in the computation of the metallicity. Cyan stars show the CALIFA-based nebular $\mu$ZR of S\'anchez et al (2013). Middle and right: The mean $\mu$ZR obtained by breaking the 300 galaxies sample into five M$_\star$ intervals, restricting the analysis to spatial regions outwards and for the inner R= 0.5 HLR }
\label{fig:fig4}
\end{figure}

\section*{Acknowledgements}
\noindent
This contribution is based on data obtained by the CALIFA survey (http://califa.caha.es), funded by the Spanish MINECO grants ICTS-2009-10, AYA2010-15081, and the CAHA operated jointly by the Max-Planck IfA and the IAA (CSIC). The CALIFA Collaboration thanks the CAHA staff for the dedication to this project. 
Support from CNPq (Brazil) through Programa Ci\^encia sem Fronteiras (401452/2012-3) is duly acknowledged.


\begin{thebibliography}{99}

\bibitem[\protect\citeauthoryear{Blanton}{2009}]{Blanton2009} Blanton, M.R., Moustakas, J., 2009, ARAA, 47, 159

\bibitem[\protect\citeauthoryear{Bundy}{2014}]{Bundy14}Bundy, K., et al.\ 2014, https://www.sdss3.org/future/manga.php

\bibitem[{{Bruzual} \& {Charlot}(2003)}]{bruzual03}{Bruzual} G., {Charlot} S., 2003, \mnras, 344, 1000

\bibitem[\protect\citeauthoryear{Cappellari}{2011}]{Cappellari2011}Cappellari, M, Emsellem, E., et al., 2011, MNRAS, 413, 813

\bibitem[\protect\citeauthoryear{Cid Fernandes}{2005}]{Cid2005} Cid Fernandes, R.,  Mateus, A.,  Sodr\'e, L.,  et al.\ 2005, \mnras, 358, 363

\bibitem[\protect\citeauthoryear{Cid Fernandes}{2013}]{Cid2013} Cid Fernandes, R., P\'erez, E., Garc\'\i a Benito, R., et al.\ 2013,  \aap, 557, 86

\bibitem[\protect\citeauthoryear{Cid Fernandes}{2014}]{Cid2014} Cid Fernandes, R., Gonz\'alez Delgado, R.M., P\'erez, E., et al.\ 2014, \aap, 561, 130

\bibitem[\protect\citeauthoryear{Croom}{2012}]{Croom14} Croom, S., Lawrence, J.S., Bland-Hawthorn, J., et al., 2012, MNRAS, 421, 872

\bibitem[\protect\citeauthoryear{Garcia-Benito}{2014}]{RGB14}Garc\'\i a-Benito, R., et al., 2014, A\&A, submitted

\bibitem[\protect\citeauthoryear{Gonz\'alez Delgado}{2005}]{RGD05} Gonz\' alez Delgado, R. M., Cervi\~no, M., Martins, et al.\ 2005, \mnras, 357, 945

\bibitem[\protect\citeauthoryear{Gonz\'alez Delgado}{2014}]{RGD14a} Gonz\' alez Delgado, R. M., P\'erez, E., Cid Fernandes, R., et al.\ 2014a, \aap, 562, 47 

\bibitem[\protect\citeauthoryear{Gonz\'alez Delgado}{2014}]{RGD14b} Gonz\' alez Delgado, R. M., Cid Fernandes, R., Garc\'\i a-Benito, R., et al.\ 2014b, \apjl, 791, L16

\bibitem[\protect\citeauthoryear{Husemann et al.}{2013}]{Husemann2013} Husemann, B., Jahnke, K., S\'anchez, S. F., et al.\ 2013, \aap, 549, A87

\bibitem[\protect\citeauthoryear{Perez}{2013}]{Perez} P\'erez, E., Cid Fernandes, R., Gonz\'alez Delgado, R. M., et al.\ 2013, \apjl, 764, 1L

\bibitem[\protect\citeauthoryear{Roberts}{1994}]{Roberts}Roberts, M.S., Haynes, M.P., ARAA, 32, 115

\bibitem[\protect\citeauthoryear{Sanchez, S.F.,  et al.}{2012}]{Sanchez2012} S\'anchez, S. F., Kennicutt, R. C., Gil de Paz, A., et al.\ 2012, \aap, 538, 8

\bibitem[\protect\citeauthoryear{Sanchez, S.F.,  et al.}{2013}]{Sanchez2013} S\'anchez, S. F., Rosales-Ortega, F.F, Jungwiert, B., et al.\ 2013, \aap, 554, 58

\bibitem[\protect\citeauthoryear{V\' azdekis, A.  et al.}{2010}]{Vazdekis2010} Vazdekis, A., S\'anchez-Bl\'azquez, P., Falc\'on-Barroso, J., et al.\ 2010, \mnras, 404, 1639

\bibitem[\protect\citeauthoryear{Walcher}{2014}]{Walcher2014} Walcher, C.J. , Wisotzki, L., Bekerait\'e, S., et al., 2014, A\&A, arXiv1407.2939

\end{thebibliography}
\end{document}